\documentclass[12pt]{article}
\pdfoutput=1
\usepackage{jhep-mod}
\usepackage{bm}
\usepackage{amssymb}
\usepackage{pifont}
\usepackage{slashed} 
\usepackage{slantsc} 

\def\d{{\mathrm{d}}}
\def\g{{\mathfrak{g}}}
\title{Explicit Baker--Campbell--Hausdorff formulae for some specific Lie algebras}
\author{Alexander Van--Brunt\, \hbox{{\sf and}} }
\emailAdd{alexandervanbrunt@gmail.com}
\author{Matt Visser\,}
\emailAdd{matt.visser@msor.vuw.ac.nz}
\affiliation{ \mbox{School of Mathematics, Statistics, and Operations Research,}\\
Victoria University of Wellington; \\
PO Box 600, Wellington 6140, New Zealand.}
\date{18 May 2015; \LaTeX-ed \today}
\abstract{
In a previous article, [arXiv:1501.02506, JPhysA {\bf48} (2015) 225207], we demonstrated that whenever $[X,Y] = u X + vY + cI$ the Baker--Campbell--Hausdorff formula
reduces to the tractable closed-form expression
\[
Z(X,Y)=\ln( e^X e^Y ) = X+Y+ f(u,v) \; [X,Y],
\]
where $f(u,v)=f(v,u)$ is explicitly given by
\[
f(u,v) = {(u-v)e^{u+v}-(ue^u-ve^v)\over u v (e^u - e^v)} =  {(u-v)-(ue^{-v}-ve^{-u})\over u v (e^{-v} - e^{-u})}.
\]
This is much more general than the results usually presented for either the Heisenberg commutator $[P,Q]=-i\hbar I$, or the creation-destruction commutator $[a,a^\dagger]=I$. In the current article we shall further generalize and extend this result, primarily by relaxing the input assumptions.  We shall work with the structure constants $f_{ab}{}^c$ of the Lie algebra, (defined by $[T_a,T_b] = f_{ab}{}^c \; T_c$), and identify suitable constraints one can place on the structure constants to make the  Baker--Campbell--Hausdorff formula tractable. We shall also develop related results using the commutator sub-algebra $[\g,\g]$ of the relevant Lie algebra~$\g$. Under suitable conditions, and taking $L_A B = [A,B]$ as usual, we shall demonstrate that
\[
\ln( e^X e^Y ) = X + Y  +  {I \over e^{-L_X} - e^{+L_Y} } \left(  {I-e^{-L_X}\over L_X} + {I-e^{+L_Y}\over L_Y} \right) [X,Y].
\]
}
\begin{document}
\maketitle
\clearpage
\section{Introduction}

The Baker--Campbell--Hausdorff formula for $Z(X,Y)=\ln( e^X e^Y )$ when $X$ and $Y$ are non-commutative quantities  is a  general multi-purpose result of considerable interest in not only both pure and applied mathematics~\cite{early-history, book, Dynkin1, Dynkin2, Goldberg, Sack, Wilcox, Newman-Thompson, Reinsch, Casas-Murua, Van-Brunt:2015a, Van-Brunt:2015b}, 
but also within the fields of theoretical physics, physical chemistry, the theory of numerical integration, and other disciplines~\cite{Van-Brunt:2015a, Van-Brunt:2015b, Fisher, Schumaker, Truax, Nieto:1993, Shanta:1993, Nieto:1996}. 
Applications include topics as apparently remote and unconnected as the embedding problem for stochastic matrices. 
For our current purposes, the general Baker--Campbell--Hausdorff formula can best be written as~\cite{Van-Brunt:2015a}
\begin{eqnarray}
\label{E:BCH}
&&\ln( e^X e^Y ) = X + Y 
+ \int_0^1 \d t  \sum_{n=1}^\infty {(I-e^{L_X} e^{tL_Y})^{n-1} \over n(n+1)} \; {( e^{L_X}-I)\over L_X} \; [X, Y].
\end{eqnarray}
Here as usual $L_A B = [A,B]$. If one makes no further simplifying assumptions, then this expression expands to an infinite series of nested commutators, with the first few well-known terms being~\cite{early-history,book}
\begin{eqnarray}
\ln( e^X e^Y ) &=& X + Y + {1\over2}\;[X,Y] + {1\over12}  L_{X-Y}[X,Y] - {1\over24} L_YL_X[X,Y] + \dots
\end{eqnarray}
Higher-order terms in the expansion quickly become very unwieldy.  
See for instance references~\cite{early-history, book, Dynkin1, Dynkin2, Goldberg, Sack, Wilcox, Newman-Thompson, Reinsch, Casas-Murua, Van-Brunt:2015a, Van-Brunt:2015b}. 
In contrast, by making specific simplifying assumptions about the commutator $[X,Y]$ one can sometimes obtain a terminating series, or develop other ways of simplifying  the expansion.
The most common terminating series results are:
\begin{itemize}
\item If $[X,Y]=0$, then: $\ln( e^X e^Y ) = X + Y$.
\item If $[X,Y]=cI$, then: $\ln( e^X e^Y ) = X + Y + {1\over2} c I$.
\item If  $[X,Y] = v Y$, then:
\begin{eqnarray}
\ln( e^X e^Y ) &=&  X + {v Y\over 1-e^{-v}} 
=  X + Y + {ve^v-e^v+1\over v(e^v-1)}\; [X,Y].
\end{eqnarray}
Observe  that $[X,Y] = v Y$ implies that $X$ acts as a ``shift operator'', a ``ladder operator'', for $Y$, thus allowing one to invoke the techniques of Sack~\cite{Sack}. 
This particular result can also be extracted from equation (7.9) of Wilcox~\cite{Wilcox}; but only after some nontrivial manipulations.
\end{itemize}
Considerably more subtle is our recent result~\cite{Van-Brunt:2015a}:
\begin{itemize}
\item 
If $[X,Y] = u X + vY + cI$, then:
\begin{equation}
\ln( e^X e^Y ) = X+Y+ f(u,v) \; [X,Y],
\end{equation}
where 
\begin{equation}
\label{E:f1}
\qquad f(u,v) = f(u,v) = {(u-v)e^{u+v}-(ue^u-ve^v)\over u v (e^u - e^v)}.
\end{equation}
It is often more useful to write this as
\begin{equation}
\label{E:f2}
f(u,v)
=  {(u-v)-(ue^{-v}-ve^{-u})\over u v (e^{-v} - e^{-u})}.
\end{equation}
Sometimes the structure is more clearly brought out by writing this in the form
\begin{equation}
\label{E:f3}
f(u,v) =  {1 \over e^{-u} - e^{-v} } \left(  {1-e^{-u}\over u} - {1-e^{-v}\over v} \right).
\end{equation}
\end{itemize}
In a series of very recent articles, Matone~\cite{Matone:2015a,Matone:2015b,Matone:2015c}, has generalized this result in various ways.\footnote{
Matone and Pasti have then also applied somewhat related ideas to the ``covariantization'' of differential operators~\cite{Matone:2015d}.}
In the current article we shall  also develop several generalizations --- but shall work towards a rather different goal by instead seeking to weaken the conditions under which this simplified form of the  Baker--Campbell--Hausdorff [BCH] formula applies.

\section{Strategy}

\noindent
In reference~\cite{Van-Brunt:2015a} our strategy was to use 
\begin{equation}
[X,Y] = u X + vY + cI,
\label{E:1}
\end{equation}
 to first deduce
\begin{equation}
L_X [X,Y] = v [X,Y]; \qquad L_Y [X,Y] = - u [X,Y].
\label{E:2}
\end{equation}
Once equation (\ref{E:2}) is established, then equation~(\ref{E:BCH}) collapses to
\begin{eqnarray}
\ln( e^X e^Y ) \to X + Y  + \int_0^1 \d t  \sum_{n=1}^\infty {(1-e^{v} e^{-tu})^{n-1} \over n(n+1)} \; {( e^{v}-1)\over v} \; [X, Y].
\end{eqnarray}
This implies
\begin{equation}
f(u,v) = {( e^{v}-1)\over v}\; \int_0^1 \d t  \sum_{n=1}^\infty {(1-e^{v} e^{-tu})^{n-1} \over n(n+1)}.
\end{equation}
Performing the sum and evaluating the integral is straightforward, (if a little tedious), with the result given in equations (\ref{E:f1})--(\ref{E:f3}). See reference~\cite{Van-Brunt:2015a} for details. 
But to obtain this final result, the key step involves the two commutators in equation (\ref{E:2}), not the original commutator in equation (\ref{E:1}). This suggests it might be more useful to focus attention on equation (\ref{E:2}), since that is a less restrictive result that does not require equation (\ref{E:1}). Are there situations where we can get equation (\ref{E:2}) to hold with  equation (\ref{E:1})
being violated?

\section{Structure constants}

Let us work in a Lie algebra with basis $T_a$, and define the structure constants $f_{ab}{}^c$ by taking $[T_a,T_b] = f_{ab}{}^c \; T_c$.
Then setting
\begin{equation}
X = x^a \,T_a \qquad \hbox{and}\qquad Y = y^a \, T_a
\end{equation}
implies
\begin{equation}
[X,Y] = (x^a \,y^b\,  f_{ab}{}^c) \; T_c.
\end{equation}
We shall now systematically build up to our most general result in several incremental stages.
The art lies in choosing structure constants appropriately.

\subsection{Case 1: Reproducing the special commutator}
Let us first choose
\begin{equation}
 f_{ab}{}^c =  m_{[a} \; \delta_{b]}{}^c.
\end{equation}
Then 
\begin{equation}
[X,Y] = {1\over2} \left\{  (x^a \,m_a) Y - (y^a \,m_a) X \right\}.
\end{equation}
If we now define
\begin{equation}
X = \hat X + \alpha I; \qquad Y = \hat Y + \beta I,
\end{equation}
then
\begin{eqnarray}
[\hat X,\hat Y] &=& {1\over2} \left\{  (x^a \,m_a) \hat Y - (y^a \,m_a) \hat X \right\} 
+{1\over2} \left\{  (x^a \, m_a) \beta - (y^a \,m_a) \alpha \right\} I.
\end{eqnarray}
This is our special commutator of equation (\ref{E:1}), under the identifications
\begin{equation}
u = - {(y^a \,m_a) \over2}; \qquad v ={(x^a \,m_a) \over2}; 
\qquad\hbox{with}\qquad 
c= {1\over2} \left\{  (x^a \,m_a) \beta - (y^a \,m_a) \alpha \right\} .
\end{equation}
Thus this particular set of structure constants has not actually generalized our previous result --- instead it has provided a natural way in which the specific commutator (\ref{E:1}) will automatically arise.

\subsection{Case 2: Commutator algebras of dimension unity}
Let us now choose
\begin{equation}
 f_{ab}{}^c =  \omega_{ab}\; n^c.
\end{equation}
Note that the special commutator of equation (\ref{E:1}) can certainly be put into this form. Specifically, by taking $T_a = (X,Y,I)$ we have
\begin{equation}
\omega_{ab} = \left[\begin{array}{rrr}0&+1&0\\-1&0&0\\0&0&0\end{array}\right]; 
\qquad\hbox{and}\qquad 
n^c = (u, v, c).
\end{equation}
But we shall now work with completely arbitrary $ n^c$ and $\omega_{ab}$, thereby generalizing our previous result.
Let us define
\begin{equation}
u= -y^a\, \omega_{ab} \,n^b; \qquad v= x^a \,\omega_{ab} \,n^b,
\end{equation}
and observe
\begin{equation}
[X,Y] = (\omega_{ab}\, x^a\, y^b)\; (n^c\, T_c).
\end{equation}
Now compute
\begin{eqnarray}
L_X [X,Y] 
&=& (\omega_{ab} \,x^a \,y^b)\; L_X (n^c \,T_c) \\
&=& (\omega_{ab} \,x^a \,y^b)\; (\omega_{ab} \,x^a \,n^b)\; (n^c \,T_c) \\
&=&  (\omega_{ab} \,x^a \,n^b)\; (\omega_{ab} \,x^a \,y^b)\; (n^c \,T_c)\\
&=& v [X,Y].
\end{eqnarray}
Similarly
\begin{eqnarray}
L_Y [X,Y] 
&=& (\omega_{ab} \,x^a \,y^b)\; L_Y (n^c \,T_c) \\
&=& (\omega_{ab} \,x^a \,y^b)\; (\omega_{ab} \,y^a \,n^b)\; (n^c \,T_c) \\
&=&  (\omega_{ab} \,y^a \, n^b)\; (\omega_{ab} \,x^a \,y^b)\; (n^c\, T_c) \\
&=& -u [X,Y].
\end{eqnarray}
This establishes equation (\ref{E:2}) \emph{without} equation (\ref{E:1}).\footnote{
More formally, this can be phrased as the statement that the commutator $[X,Y]$ be a simultaneous eigenvector of the two adjoint operators $L_X$ and $L_Y$.}   Consequently
\begin{equation}
\ln(e^X e^Y)  = X+Y+f(u,v)[X,Y],
\end{equation}
for the \emph{same} function $f(u,v)$ as previously encountered.
More explicitly we now have
\begin{eqnarray}
\label{E:explicit}
\ln(e^X e^Y)  &=& X+Y
+f(x^a \,\omega_{ab} \, n^b,-y^a \,\omega_{ab}\, n^b) \;\; (\omega_{ab}\, x^a \, y^b)\;\; (n^c \,T_c).
\end{eqnarray}
We can also write this as
\begin{eqnarray}
&&\ln(e^Xe^Y) = 
\left\{ x^c+ y^c + f(x^a\, \omega_{ab} \,n^b,-y^a \,\omega_{ab} \, n^b) \; (\omega_{ab} \,x^a \,y^b)\; n^c \right)\} \;\;T_c.
\end{eqnarray}
That is, (at least in this particular class of Lie algebras), we see that we can view the Baker--Campbell--Hausdorff formula as a generalized notion of ``addition''. 
By defining the generalized ``addition'' operator $\oplus$ via $  (x\oplus y)^c \;T_c = \ln(e^Xe^Y)$, we explicitly have
\begin{eqnarray}
&& (x\oplus y)^c =
x^c+ y^c + f(x^a \omega_{ab} n^b,-y^a \,\omega_{ab} \, n^b) \; (\omega_{ab} \,x^a \, y^b)\; n^c.
\end{eqnarray}
Now the statement that $f_{ab}{}^c = \omega_{ab} \; n^c$ can be rephrased as  the statement that the commutator sub-algebra $[\g,\g]$, (the sub-algebra formed from the commutators of the ambient  Lie algebra $\g$), is of dimension unity.\footnote{ We typically take $\g$ to be some arbitrary but fixed ambient Lie algebra, with both $X\in \g$ and $Y\in\g$.  Alternatively we might initially take $\g$ to be the minimal free Lie algebra generated by $X$ and $Y$, but then might add some constraints (eg: nilpotency, solvability, etc) to modify that free algebra.} 
Note that the object $[\g,\g]$ is also called the first derived sub-algebra, or the first lower central sub-algebra, 
(\emph{aka} first descending central sub-algebra), 
though these two series of sub-algebras will differ once one goes to higher levels.
%
Observe  that:
\begin{itemize}
\item 
If the commutator sub-algebra $[\g,\g]$ is of dimension zero, then the Lie algebra is Abelian, and the Baker--Campbell--Hausdorff result 
    is trivial: $\ln(e^Xe^Y) = X+Y$. 
\item 
If the commutator sub-algebra  $[\g,\g]$ is of dimension one then $[T_a,T_b] \propto N$, for some fixed $N$.
Now write $N = n^c\, T_c$, then $[T_a,T_b] \propto (n^c\, T_c)$, thereby implying $[T_a,T_b] = \omega_{ab} \,n^c \,T_c$.
\item 
We can naturally split this into 2 sub-cases:
\begin{equation}
\omega_{ab} \,n^b = 0 \qquad \hbox{and}  \qquad \omega_{ab} \,n^b \neq 0.
\end{equation}
\item 
If $\omega_{ab} \,n^b = 0$, then both 
\begin{equation}
u= y^a \,\omega_{ab} \, n^b=0; \qquad \hbox{and} \qquad v=-x^a \,\omega_{ab}\, n^b = 0.
\end{equation}
Therefore $L_X[X,Y]=0=L_Y[X,Y]$, and so
\begin{equation}
[\g,[\g,\g]]=0.
\end{equation}
That is, the second lower central sub-algebra is trivial, and in particular the original Lie algebra is nilpotent.
    (Examples: The Heisenberg algebra and the creation-destruction algebra.) 
\item
    If  $\omega_{ab} \, n^b \neq 0$ then $u$ and $v$ are nontrivial, and $f(u,v)$ is also nontrivial.
    The Lie algebra is now not nilpotent but satisfies the more subtle condition that
\begin{equation}
[\g,[\g,\g]] =[\g,\g].
\end{equation}
That is, the second lower central sub-algebra, (and so all the higher-order lower central sub-algebras), all equal the first lower central sub-algebra. 
This can also be phrased as the demand that the commutator sub-algebra be an ideal of the underlying Lie algebra.
\end{itemize}
In short, the explicit Baker--Campbell--Hausdorff formula (\ref{E:explicit}) holds whenever the commutator sub-algebra  $[\g,\g]$ is of dimension unity.

\bigskip
\subsection{Case 3: Nilpotent Lie algebras}

Consider now the higher terms in the lower central series,  defined iteratively by
\begin{equation}
\g_0  =\g;  \qquad \g_1=[\g,\g];   \qquad \g_n = [\g, \g_{n-1}].
\end{equation}
If, for some $n$, we have $\g_n=0$ then the Lie algebra $\g$ is said to be ``nilpotent''. In this case all $n$th-order and higher commutators vanish and the Baker--Campbell--Hausdorff series truncates --- but this result has previously been (implicitly) used when developing the Reinsch algorithm~\cite{Reinsch, Casas-Murua}, and our own variant thereof~\cite{Van-Brunt:2015b}. That algorithm works by utilizing a faithful representation for the first $n$ nested commutators of a free Lie algebra in terms of strictly upper triangular $(n+1)\times(n+1)$ matrices with entires only on the first super-diagonal. That is: working with a level-$n$ nilpotent Lie algebra ``merely'' reproduces the first $n$ terms in the Baker--Campbell--Hausdorff formula, and gives zeros thereafter. So while certainly useful, this is not really new~\cite{Reinsch, Casas-Murua,Van-Brunt:2015b}.
In terms of the structure constants, nilpotency is achieved if at some stage
\begin{equation}
f_{ab}{}^i \; f_{ic}{}^j \; f_{jd}{}^k \; f_{ke}{}^m \dots = 0.
\end{equation}

\subsection{Case 4: Abelian commutator algebras}

Can the discussion above be generalized even further? Note that in all generality
\begin{equation}
[[T_a,T_b],[T_c,T_d]] = f_{ab}{}^m f_{cd}{}^n f_{mn}{}^e\;T_e.
\end{equation}
So whenever $f_{ab}{}^c = \omega_{ab}\; n^c$, we have
\begin{equation}
[[T_a,T_b],[T_c,T_d]] = 0,
\end{equation}
or more abstractly
\begin{equation}
[[\g,\g],[\g,\g]] = 0.
\end{equation}
That is,  the commutator sub-algebra is Abelian.
This is a special case of a ``solvable'' Lie algebra. 
Can anything be done for more general  solvable Lie algebras?

Let us now consider the situation where the the commutator sub-algebra is Abelian, but we do not demand that the commutator algebra is one dimensional. The Jacobi identity leads to
\begin{eqnarray}
L_X L_Y W &=& [X,[Y,W]]
\\
&=& - [Y,[W,X]] - [W,[X,Y]]
\\
&= & [Y,[X,W]] + [[X,Y],W]
\\
&=& L_Y L_X W + L_{[X,Y]} W.
\end{eqnarray}
That is
\begin{equation}
\label{E:Jacobi-eff}
[L_X,L_Y] W =  L_{[X,Y]} W.
\end{equation}
But if $W$ is itself a commutator, $W=[U,V]$, and if the commutator algebra is Abelain, $[[\g,\g],[\g,\g]]=0$, then we have
\begin{equation}
[L_X,L_Y] [U,V] = 0.
\end{equation}
That is, in this situation, and \emph{when acting on commutators}, $L_X$ and $L_Y$ commute. But this is exactly the situation in the Baker--Campbell--Hausdorff expansion of equation (\ref{E:BCH}), $L_X$ and $L_Y$ are always acting on commutators. So as long as the commutator sub-algebra is itself Abelian we can rearrange equation (\ref{E:BCH}) to write
\begin{eqnarray}
\label{E:BCH2}
&&\ln( e^X e^Y ) = X + Y 
 +{( e^{L_X}-I)\over L_X}  \int_0^1 \d t  \sum_{n=1}^\infty {(I-e^{L_X} e^{tL_Y})^{n-1} \over n(n+1)} \; [X, Y],
\end{eqnarray}
and treat the $L_X$ and $L_Y$ \emph{as though they commute}. 
But then, summing and integrating as previously, we have
\begin{equation}
\label{E:BCH3}
\ln( e^X e^Y ) = X + Y + f(L_X, -L_Y) [X,Y],
\end{equation}
for exactly the same function $f(u,v)$, subject only to the condition $[[\g,\g],[\g,\g]]=0$ that the commutator algebra be Abelian.
Note that this last formula is still an operator equation, which still contains an infinite set of nested commutators --- albeit in a relatively explicit manner. 
Indeed, under the stated conditions, from equation (\ref{E:f3}) we see\footnote{ 
With considerable hindsight, reinterpretation, and rearrangement, this result can be seen to be closely related to the meta-Abelian analysis of Kurlin~\cite{Kurlin}.}
\begin{eqnarray}
\label{E:BCHf}
&&\ln( e^X e^Y ) = X + Y
 +  {I \over e^{-L_X} - e^{+L_Y} } \left(  {I-e^{-L_X}\over L_X} + {I-e^{+L_Y}\over L_Y} \right) [X,Y].
\end{eqnarray}
Note that in terms of the structure constants the condition $[[\g,\g],[\g,\g]]=0$ is equivalent to the explicit constraint
\begin{equation}
f_{ab}{}^m \;f_{cd}{}^n \; f_{mn}{}^e= 0.
\end{equation}
Furthermore, we note the series expansion
\begin{eqnarray}
\label{E:series}
f(u,v) &=& {1\over2} +{u+v\over12}+ {uv\over24}
- {(u+v)(u^2-5uv+v^2)\over720}
-{uv(u^2-4uv+v^2)\over1440} + \dots\qquad\quad
\end{eqnarray}
which verifies that, (as it should), the operator $f(L_X, -L_Y)$ contains only non-negative powers of $L_X$ and $L_Y$.

\subsection{Case 5: $[X,Y]$ is in the centre of the commutator algebra}

Let us now relax the conditions for the validity of this result even further: The key step is to realize that
\begin{equation}
[L_X,L_Y] [U,V] =  L_{[X,Y]} [U,V],
\end{equation}
so that $L_X$ (effectively) commutes with $L_Y$ as long as $[[X,Y],[U,V]] = 0$. 
As previously noted, this certainly holds as long as the commutator algebra is Abelian, $[[\g,\g],[\g,\g]] = 0$, but it is quite sufficient to demand
that the \emph{specific} commutator $[X,Y]$ is an element of the centre $Z_{[\g,\g]}$ of the commutator algebra $[\g,\g]$. That is
\begin{equation}
[[X,Y],[\g,\g]]=0.
\end{equation}
In terms of the structure constants this is equivalent to the weakened constraint
\begin{equation}
x^a\;y^b\; f_{ab}{}^m\; f_{cd}{}^n\; f_{mn}{}^e= 0.
\end{equation}

\noindent
Under this milder condition we still have (effective) commutativity of  $L_X$  with $L_Y$,  thereby allowing us to treat the $L_X$  and $L_Y$  appearing in the Baker--Campbell--Hausdorff formula (\ref{E:BCH}) \emph{as though} they commute. Integrating and summing the series we again see
\begin{equation}
\label{E:BCH3}
\ln( e^X e^Y ) = X + Y + f(L_X, -L_Y) [X,Y],
\end{equation}
again for exactly the same function $f(u,v)$, now subject only to the weaker condition that $[[X,Y],[\g,\g]]=0$. Under the stated conditions, that the \emph{specific} commutator $[X,Y]$ be an element of the centre of the commutator algebra $[\g,\g]$, we again find
\begin{eqnarray}
\label{E:BCHf-case5}
&&\ln( e^X e^Y ) = X + Y
 +  {I \over e^{-L_X} - e^{+L_Y} } \left(  {I-e^{-L_X}\over L_X} + {I-e^{+L_Y}\over L_Y} \right) [X,Y].
\end{eqnarray}

\medskip
\subsection{Case 6: $[X,Y]$ is in the centralizer of $\{L_X^m L_Y^n [X,Y]\}$}
\def\S{{\mathcal{S}}}

As our final weakening of the input assumptions, (while still keeping the same strength conclusions), take an arbitrary but fixed ambient Lie algebra $\g$ and consider the set
\begin{equation}
\S = \{ L_X^m L_Y^n [X,Y];  \; m\geq 0, n\geq 0\}. 
\end{equation}
The construction of this set is inspired by considering the form of the terms which appear in the Baker--Campbell--Hausdorff expansion (\ref{E:BCH}).\footnote{
If in contrast we were to take $\g$ as the minimal free algebra generated by $X$ and $Y$, then this would not be a weakening of case 5, it would merely be a restatement of case 5.}

If we now demand merely that $[X,Y]$ commute with all the elements of $\S$, (that is, $[[X,Y],\S]=0$ or equivalently $L_{[X,Y]} \S =0$, so that $[X,Y]$ is in the so-called centralizer of the set $\S$), then the Jacobi identity, (in the form (\ref{E:Jacobi-eff})), implies 
\begin{equation}
[L_X,L_Y] L_X^m L_Y^n [X,Y] = 0,
\end{equation}
which we could also write as
\begin{equation}
[L_X,L_Y] \S = 0.
\end{equation}
Then  in particular
\begin{eqnarray}
L_Y  L_X^m L_Y^n [X,Y] &=& L_X L_Y L_X^{m-1} L_Y^n[X,Y]\nonumber\\
&=& L_X^2 L_Y L_X^{m-2} L_Y^n[X,Y]\nonumber\\
&=& \dots\nonumber\\
&=& L_X^m L_Y^{n+1} [X,Y].
\end{eqnarray}
That is, under these conditions $L_X$ and $L_Y$ can still be treated \emph{as though} they commute in the Baker--Campbell--Hausdorff  expansion. 
Under these conditions,  all of the terms appearing in the Baker--Campbell--Hausdorff expansion of equation~(\ref{E:BCH}) can now be reduced to elements of the set $\S$. 
Integrating and summing the series we again see
\begin{equation}
\label{E:BCH4}
\ln( e^X e^Y ) = X + Y + f(L_X, -L_Y) [X,Y],
\end{equation}
again for exactly the same function $f(u,v)$, but now subject only to the even weaker condition $[[X,Y],\S]=0$, that the specific commutator $[X,Y]$ be an element of the centralizer of $\S$. 
To be explicit about this, under the stated conditions
\begin{eqnarray}
\label{E:BCHf-case6}
&&\ln( e^X e^Y ) = X + Y
 +  {I \over e^{-L_X} - e^{+L_Y} } \left(  {I-e^{-L_X}\over L_X} + {I-e^{+L_Y}\over L_Y} \right) [X,Y].
\end{eqnarray}
Careful inspection of the above quickly verifies that the \emph{only} terms present when one expands the above are of the form $L_X^m L_Y^{n} [X,Y]$, (the elements of the set $\S$), and that our simplifying assumption has eliminated all terms such as $L_{[X,Y]} L_X^m L_Y^{n} [X,Y]$, and variants thereof. By summing over the integers $m$ and $n$ the centralizer condition can also be restated as
\begin{equation}
[ [X,Y],  e^{sL_X} e^{tL_Y} [X,Y]] = 0;  \qquad\qquad \forall s, t.
\end{equation}
This is as far as we have currently been able to weaken the input assumptions we originally started with, while still keeping a reasonably close analogue of our initial result involving the function $f(u,v)$.

\clearpage
\section{Discussion}

The Baker--Campbell--Hausdorff formula is a general purpose tool that has found many applications both in pure and applied mathematics~\cite{early-history, book, Dynkin1, Dynkin2, Goldberg, Sack, Wilcox, Newman-Thompson, Reinsch, Casas-Murua, Van-Brunt:2015a, Van-Brunt:2015b}, and generally in the physical sciences~\cite{Van-Brunt:2015a, Van-Brunt:2015b, Fisher, Schumaker, Truax, Nieto:1993, Shanta:1993, Nieto:1996}.  Via the study of the embeddability problem for stochastic matrices (Markov processes) there are even potential applications in the social sciences and financial sector. Explicit closed-form results are relatively rare, see the Introduction for examples. 
In this present article we have significantly extended our previous results reported in reference~\cite{Van-Brunt:2015a} by systematically weakening the input assumptions. 
In a number of increasingly general situations we have shown that the Baker--Campbell--Hausdorff expansion can be written in closed form as 
\begin{equation}
\label{E:BCH5}
\ln( e^X e^Y ) = X + Y + f(u,v) [X,Y],
\end{equation}
where $f(u,v)$ is the symmetric function
\begin{equation}
\label{E:ff}
f(u,v) = f(v,u) = {(u-v)e^{u+v}-(ue^u-ve^v)\over u v (e^u - e^v)}.
\end{equation}
This was first demonstrated in reference~\cite{Van-Brunt:2015a} for the very explicit commutator $[X,Y]=uX+vY+cI$.  Herein, (with suitable expressions for $u$ and $v$),  a structurally  identical result is established for Lie algebras with a one-dimensional commutator sub-algebra. 
More generally, whenever the commutator sub-algebra is Abelian, one has
\begin{equation}
\label{E:BCH6}
\ln( e^X e^Y ) = X + Y + f(L_X, -L_Y) [X,Y].
\end{equation}
More specifically
\begin{equation}
\ln( e^X e^Y ) = X + Y +  {I \over e^{-L_X} - e^{+L_Y} } \left(  {I-e^{-L_X}\over L_X} + {I-e^{+L_Y}\over L_Y} \right) [X,Y].
\end{equation}
This result furthermore  extends to the weaker input condition $[X,Y]\in Z_{[g,g]}$, that is, $[X,Y]$ being an element of the centre of the commutator algebra. Even more generally, this result extends to $[X,Y]$ being an element of the centralizer of  those Lie brackets that appear in the Baker--Campbell--Hausdorff expansion. 
Overall, we find it quite remarkable just how far we have been able to push this result. There are of course many other directions that one might also wish to explore --- we have concentrated our efforts on directions in which it seems that relatively concrete and explicit results might be readily extractable.

\acknowledgments

This research was supported by the Marsden Fund, through a grant administered by the Royal Society of New Zealand. 
AVB was also supported by a Victoria University of Wellington Summer Scholarship.


\end{document}